\documentclass[11pt]{article}
\usepackage{fullpage, times, amsmath}
\newcommand{\defeq}{\stackrel{\Delta}{=}}
\newcommand{\ket}[1]{| #1 \rangle}

\newcommand{\braket}[2]{\langle #1 | #2 \rangle}
\newcommand{\braopket}[3]{\langle #1 | #2 | #3 \rangle}

\newcommand{\rn}{\mathrm{rn}}
\newcommand{\cn}{\mathrm{cn}}
\newcommand{\rw}{\mathrm{rw}}
\newcommand{\cw}{\mathrm{cw}}
\newcommand{\rcw}{\mathrm{rcw}}
\newcommand{\trnorm}[1]{|| #1 ||_{\mathrm{tr}}}
\newcommand{\frnorm}[1]{|| #1 ||_{\mathrm{F}}}
\newcommand{\convw}{\mathrm{convw}}
\newcommand{\rank}{\mathrm{rank}}
\newcommand{\junk}{\mathrm{junk}}
\newcommand{\norm}[1]{\| #1 \|}
\newcommand{\Qsim}{Q^{\|}}
\newcommand{\Rsimpub}{R^{\|, \mathrm{pub}}}

\newtheorem{definition}{Definition} 
\newtheorem{fact}{Fact}    
\newenvironment{proof}{\noindent{\bf Proof:}}{\qed} 
\newcommand{\qed}{\hfill{$\rule{6pt}{6pt}$}} 

\title{A note on the power of quantum fingerprinting}
\author{
Alexander Golynski\thanks{
School of Computer Science,
University of Waterloo, 
Waterloo, Ontario N2L 3G1, Canada. 
Email:  {\sf agolynsk@uwaterloo.ca}.
}
\and
Pranab Sen\thanks{
NEC Laboratories America,
4 Independence Way, Suite 200,
Princeton, NJ 08540, U.S.A.
Email:  {\sf pranab@nec-labs.com}.
Work done while the author was a postdoctoral researcher at the
University of Waterloo.
}
}
\date{}

\begin{document}
\maketitle

\begin{abstract}
In this short note, we improve and extend Yao's paper "On the
power of quantum fingerprinting" \cite{Yao03} about
simulating a classical public coin simultaneous message protocol 
by a quantum simultaneous message protocol with no shared resource.
\end{abstract}

\section{Introduction}
The {\em simultaneous message} model of communication complexity can
be described as follows. Suppose 
$f: \{0, 1\}^n \times \{0, 1\}^n \rightarrow \{0, 1\}$ 
is a function. There are three players viz. Alice, Bob
and a referee. Alice possesses $x \in \{0, 1\}^n$ and 
Bob possesses $y \in \{0, 1\}^n$.
Alice and Bob each send a single message to the referee, who then
outputs a guess for $f(x, y)$.
Alice's and Bob's messages can be classical or quantum.

In the {\em classical public coin simultaneous message} model,
Alice, Bob and the referee know the state of an additional
random variable called a {\em public coin},
that is chosen independently of the input $(x, y)$ according
to some probability distribution.
The messages $a$ and $b$ of Alice and Bob respectively
are deterministic functions of the
state $l$ of the public coin as well as the input $(x, y)$
i.e. $a = a(x,l)$ and $b = b(y,l)$. Suppose $a$ is at most $c_A$
bits long and $b$ is at most $c_B$ bits long, for any input 
$(x, y)$ and any value $l$ of the public
coin. Given the state $l$ of the public coin,
the strategy of the referee is deterministic and 
can be modeled by an $M_A \times M_B$ boolean matrix $D_l$, where
$M_A \defeq 2^{c_A}$, $M_B \defeq 2^{c_B}$. 
The rows and columns of $D_l$ are indexed by
the possible messages of Alice and Bob respectively.
On receiving messages $a$ and $b$ from Alice and
Bob respectively, the referee outputs $D_l(a, b)$.
We require that the protocol be correct with probability at least 
$3/4$ for all inputs, that is,
\[
\forall x,y \in \{0,1\}^n: ~ 
\Pr_l [D_l(a(x,l), b(y,l)) = f(x,y)] > 3/4,
\]
where the probability is over the choice of the public coin $l$.
By a result of Newman~\cite{N91}, one can assume that the public
coin $l$ is chosen uniformly from the set $[L]$, where $L = O(n)$, at
the expense of making the correctness probability at least $2/3$.
The {\em communication cost} of the protocol is defined to be 
$c_A + c_B$.
We let $\Rsimpub(f)$ denote the communication complexity of
$f$ in this model i.e. the smallest communication cost of a protocol
in this model computing $f$.

In the {\em quantum simultaneous message} model, Alice, Bob and the
referee are quantum computers. There is no prior entanglement 
amongst them i.e. at the start of the protocol, the states of
Alice, Bob and the referee are in tensor with each other.
Alice sends a pure state $\ket{u_x}$ on $c_A$ qubits and Bob sends a 
pure state $\ket{v_y}$ on $c_B$ qubits to the referee. The 
states $\ket{u_x}$ and
$\ket{v_y}$ are called the fingerprints of Alice's and Bob's
inputs $x$ and $y$ respectively. The referee performs a 
two-outcome POVM on $\ket{u_x} \otimes \ket{v_y}$ and
outputs the result. 
We require that the protocol be correct with probability at least 
$3/4$ for all inputs. 
The {\em communication cost} of the protocol is defined to be 
$c_A + c_B$.
We let $\Qsim(f)$ denote the communication complexity of
$f$ in this model.

In a recent paper, Yao~\cite{Yao03} showed how to simulate a 
classical public coin simultaneous protocol by a quantum simultaneous
protocol that has no prior entanglement. The simulation incurs an
exponential overhead. More precisely, he showed that
$\Qsim(f) \leq O(2^{2 \Rsimpub(f)} (\Rsimpub(f) + \log n))$. 
He also defined a quantity called the {\em convex width} 
$\convw(D)$ of an $M \times M$ matrix $D$, and remarked that
$\convw(D) \leq M$ if all entries of $D$ are either $0$ or $1$. 
If $D$ is the referee matrix of
an optimal classical public coin simultaneous message protocol 
for $f$ (note that in Yao's paper the referee matrix is 
assumed to be square and independent of the public coin), Yao showed 
that, in fact, $\Qsim(f) \leq O(\convw(D)^4 (\Rsimpub(f) + \log n))$.

In this note, we strengthen and generalise Yao's results.
We start by proving 
a near quadratic improvement of Yao's general 
simulation of a classical public coin simultaneous message protocol 
by a quantum simultaneous message protocol without prior entanglement. 
More precisely, we show that
$\Qsim(f) \leq O(2^{\Rsimpub(f)} (\Rsimpub(f) + \log n + 1))$. 
The same result was independently obtained 
by Gavinsky, Kempe and de Wolf~\cite{gkw:sim}. A similar
result for a related setting of communication complexity was
recently proved by Gavinsky~\cite{gavinsky:sim}. We then 
define a new notion called the {\em row-column width} $\rcw(D)$ of  
an $M \times M$ matrix $D$. For the row-column width, we can assume
that our matrix is square without loss of generality, since a 
non-square matrix can be made square by padding with zeroes without
changing its row-column width. For all square matrices $D$, 
$\rcw(D) \leq \convw(D)$. Also for $M \times M$ matrices $D$ with
boolean entries, $\rcw(D) \leq \sqrt{M}$. We show that 
if $D$ is the referee matrix of
an optimal classical public coin simultaneous message protocol 
for $f$, $\Qsim(f) \leq O(\convw(D)^4 (\Rsimpub(f) + \log n + 1))$.

The notation $\norm{\cdot}$ below stands for the $\ell_2$-norm
of a vector.

\section{An almost quadratic improvement of Yao's general simulation}
Consider an optimal classical public coin simultaneous message 
protocol for $f$. Let $c_A$ and $c_B$ be upper bounds on the message 
lengths of Alice and Bob respectively for any input $(x, y)$ and any 
value $l$ of the public coin. Define $M_A \defeq 2^{c_A}$,
$M_B \defeq 2^{c_B}$. Assume without loss of generality that
$c_A \leq c_B$.  Then, $M_A \leq 2^{\Rsimpub(f)/2}$. 
Define 
\[
\ket{u_x} \defeq \frac{1}{\sqrt{L}}
\sum_{l \in L} \ket{l} \otimes \ket{a(x,l)},
~~~
\ket{v_y} \defeq \frac{1}{\sqrt{L M_A}}
\sum_{l \in L} \ket{l} \otimes D_l \ket{b(y,l)},
\]
where matrix $D_l$ is considered as a linear operator from Bob's
message space to Alice's message space 
viz. $D_l \ket{b} \defeq \sum_{a} D_l(a,b) \ket{a}$.
Now the inner product
\[
\braket{u_x}{v_y} = \frac{1}{\sqrt{M_A}}
\sum_{l \in L} \frac{\braopket{a(x,l)}{D_l}{b(y,l)}}{L},
\]
that is, $\braket{u_x}{v_y}$ is the probability that the classical 
protocol outputs $1$ divided by $\sqrt{M_A}$.
$\norm{\ket{u_x}} = 1$. Since each $D_l$ is an $M_A \times M_B$ boolean
matrix, $\norm{D_l \ket{b}} \leq \sqrt{M_A}$ for all possible
classical messages $\ket{b}$ of Bob. Hence, $\norm{\ket{v_y}} \leq 1$. 

Now define
\[
\ket{\hat{u}_x} \defeq \ket{0}\ket{u_x},
~~~
\ket{\hat{v}_y} \defeq \ket{0}\ket{v_y} + \ket{1}\ket{\junk_y},
\]
where $\ket{\junk_y}$ 
is added to ensure that $\norm{\ket{\hat{v}_y}} = 1$.
Hence, $\braket{\hat{u}_x}{\hat{v}_y} = \braket{u_x}{v_y}$.
Thus if $f(x, y) = 1$, 
$\braket{\hat{u}_x}{\hat{v}_y} \geq (3/4) \cdot \sqrt{M_A}$, and
if $f(x, y) = 0$, 
$\braket{\hat{u}_x}{\hat{v}_y} \leq (1/4) \cdot \sqrt{M_A}$.
By a Chernoff bound (see e.g.~\cite[Corollary A.1.7]{AS00}), 
Alice and Bob can each send $O(M_A^2)$ independent
copies of $\ket{\hat{u}_x}$ and $\ket{\hat{v}_y}$ respectively
to the referee, who can then determine $f(x, y)$ with error
probability at most $1/4$ by inner product estimation via the
controlled-swap circuit~\cite{BCWW01, Yao03}. 
It follows that
$\Qsim(f) \leq O(2^{\Rsimpub(f)} (\Rsimpub(f) + \log n + 1))$. 

\section{Fingerprinting and row-column width}
We now generalize the construction of the above section, and
in the process, also generalize Yao's convex width~\cite{Yao03} to
get our new notion of row-column width. 
Define $M \defeq \max\{M_A, M_B\}$. For convenience of notation,
we assume that the possible
classical messages of Alice as well as Bob come from the set $[M]$.
Thus, we assume that for any state $l$ of the public coin
the referee matrix $D_l$ is an $M \times M$ matrix. This assumption
is without loss of generality, as will become clear later.

For an $M \times N$ matrix $Q$, define the {\em column norm} of
$Q$ as 
\[
\cn(Q) \defeq \max_{b \in [N]} \norm{Q_b},
\]
where $Q_b$ denotes the $b$th column of $Q$. The {\em row norm}
of $Q$, $\rn(Q)$, is defined similarly.
Fix an integer $K > 0$.
For every $l \in [L]$, decompose $D_l$ as a product
$D_l = E_l F_l$ for some $M \times K$ matrix $E_l$ and 
$K \times M$ matrix $F_l$. 
Define the {\em row width}, {\em column width} and 
{\em row-column width} of $D \defeq \{D_l\}_{l \in L}$ according to
the above decompositions as follows.
\begin{eqnarray*}
\rw(D)  & \defeq & 
\sqrt{\frac{1}{L} \sum_{l \in [L]} \rn(E_l)^2}, \\
\cw(D)  & \defeq & 
\sqrt{\frac{1}{L} \sum_{l \in [L]} \cn(F_l)^2}, \\
\rcw(D) & \defeq & 
\rw(D) \cdot \cw(D).
\end{eqnarray*}
The row-column width of $D \defeq \{D_l\}_{l \in L}$ is defined to
be the minimum row-column width over all decompositions of $D_l$
into products of $M \times K$ and $K \times M$ matrices, where
$K = M^2$.

Fix such optimal decompositions of $D_l$ with $K = M^2$.
Define 
\[
\ket{u_x} \defeq \frac{1}{\rw(D) \sqrt{L}}
\sum_{l \in L} \ket{l} \otimes E_l^\dagger \ket{a(x,l)}, 
~~~
\ket{v_y} \defeq \frac{1}{\cw(D) \sqrt{L}}
\sum_{l \in L} \ket{l} \otimes F_l \ket{b(y,l)}.
\]
It is easy to check that $\norm{\ket{u_x}} \leq 1$ and
$\norm{\ket{v_y}} \leq 1$. 
Now the inner product
\[
\braket{u_x}{v_y} = \frac{1}{\rcw(D)} \sum_{l \in [L]} 
                    \frac{\braopket{a(x,l)}{D_l}{\ket{b(y,l)}}}{L},
\]
that is, $\braket{u_x}{v_y}$ is the probability that the classical 
protocol outputs $1$ divided by $\rcw(D)$.

Now define
\[
\ket{\hat{u}_x} \defeq \ket{00}\ket{u_x} + \ket{01}\ket{\junk_x},
~~~
\ket{\hat{v}_y} \defeq \ket{00}\ket{v_y} + \ket{10}\ket{\junk'_y},
\]
where 
$\ket{\junk_x}$, $\ket{\junk'_y}$ are added to ensure that 
$\norm{\ket{\hat{u}_x}} = 1$, $\norm{\ket{\hat{v}_y}} = 1$ 
respectively.
Hence, $\braket{\hat{u}_x}{\hat{v}_y} = \braket{u_x}{v_y}$.
Reasoning as in the previous section, we get that
$\Qsim(f) \leq O(rcw(D)^4 (\Rsimpub(f) + \log n + 1))$. 

\section{Two properties of the row-column width}
For $l \in [L]$, consider the trivial decomposition
$D_l = I D_l$, where $I$ denotes the $M \times M$ identity matrix.
Since each $D_l$ is an $M \times M$ boolean matrix, 
\[
\cw(D) \leq \sqrt{\frac{1}{L} \sum_{l \in [L]} \cn(D_l)^2} 
       \leq \sqrt{M}
\]
for this decomposition. Also for this decomposition $\rw(D) = 1$.
Thus, for $M \times M$ boolean matrices 
the row-column width $\rcw(D) \leq \sqrt{M}$.

Consider now the case when all the $D_l$'s are the same. We shall
denote them by $D$. $D$ is an $M \times M$ boolean matrix.
We will show below that $\rcw(D) \leq \convw(D)$,
where $\convw(D)$ is the {\em convex width} of $D$ defined by
Yao~\cite{Yao03}. We first recall the definition of $\convw(D)$.
\begin{definition}[\cite{Yao03}]
$\convw(D)$ is the minimum integer $W$ for which there
exists a decomposition $D = \sum_{j=1}^W G_j P_j$,
where each $P_j$ is an $M \times M$ permutation matrix and each
$G_j$ is a symmetric positive semidefinite matrix with
non-negative real entries.
\end{definition}
Yao~\cite{Yao03} also remarked that $\convw(D) \leq M$
for any $M \times M$ boolean matrix $D$.
Indeed, consider the following cyclic diagonal decomposition 
$D = \sum_{j=1}^M D_j$, where 
\[
\begin{array}{lcll}
D_j(a,b) & \defeq & D(a,b) & {\rm if} ~ b - a \equiv (j-1) \bmod M \\
         & \defeq & 0      & {\rm otherwise}.
\end{array}
\]
Above $1 \leq a, b \leq M$.
Note that $D_j$ can be obtained by permuting the columns of a 
diagonal matrix with boolean entries. 
This decomposition shows that $\convw(D) \leq M$. In fact,
the upper bound can be attained. Consider, for example, the matrix
$Q$ where the first column is filled with all $1$'s and all other
entries are $0$. Any decomposition of $Q$ as a sum of $W < M$
matrices with non-negative real entries must contain a matrix
$Q_j$ with at least two non-zero entries in the first column and all
zeroes in the remaining columns. No permutation of the columns of
$Q_j$ can make it symmetric. This shows that $\convw(Q) = M$.
Note however that for this example, $\rcw(Q) \leq 1$.

Consider an optimal decomposition $D = \sum_{j=1}^W G_j P_j$, where
$W = \convw(D)$. Write each $G_j$ as $G_j = T_j^\dagger T_j$, where
$T_j$ is a matrix with real entries. Let $E_j \defeq T_j^\dagger$
and $F_j \defeq T_j P_j$. Let $K \defeq M W$. Since $W \leq M$,
$K \leq M^2$.
Define the $M \times K$ matrix $E$ as $E \defeq [E_1 | \ldots | E_W]$ 
and the $K \times M$ matrix $F$ as 
$F^\dagger \defeq [F_1^\dagger | \ldots | F_W^\dagger]$.
Then $D = E F$. For this decomposition of $D$ 
it is easy to see that
\[
\rn(E_j) = \cn(F_j) = \sqrt{\max_{(a,b) \in [M] \times [M]} G_j(a,b)}.
\]
Since for all $(a,b) \in [M] \times [M]$, $0 \leq G_j(a,b) \leq 1$,
$\rn(E_j) = \cn(F_j) \leq 1$, $1 \leq j \leq W$. Hence,
\[
\begin{array}{lclcl}
\rn(E) & \leq & \sqrt{\sum_{j=1}^W \rn(E_j)^2} & \leq & \sqrt{W}, \\
\cn(F) & \leq & \sqrt{\sum_{j=1}^K \cn(F_j)^2} & \leq & \sqrt{W}.
\end{array}
\]
This proves that $\rcw(D) \leq \convw(D)$.

We make two more easy observations. The first one is that
$\rcw(D) \leq \cn(D) \leq \norm{D}$,
where $\norm{D}$ denotes the $\ell_2$-operator norm of the matrix
$D$. The second one is that $\rcw(D) \leq \rank(D)$.

\begin{fact}[\cite{ronald:priv}]
\label{fact:ronald}
$O(\sqrt{M})$ is the best possible upper bound for the row-column width
of  a general $M \times M$ boolean matrix $D$.
\end{fact}
\begin{proof}
For a matrix A, we consider two norms:
\begin{eqnarray*}
\trnorm{A} & \defeq & \text{ is the sum of singular values of $A$ (trace
  norm)} \\
\frnorm{A} & \defeq & \sqrt{ \sum_{ij} \norm{A_{ij}}^2 } \text{ is the
  Frobenius norm}
\end{eqnarray*}
Let $M = 2 n$, and let $D$ be the Boolean $M \times M$ matrix for inner
product on $n$-bit strings i.e. $D_{xy} = x \cdot y \mod 2$. Let
$D_{\pm} = 2D - J$, where 
$J$ is the all-ones matrix. $J$ has rank 1 and 
$\trnorm{J} = M$. Since $D_{\pm}$ is the unnormalized $n$-qubit Hadamard
transform, we have $D_{\pm}^2 = M \cdot I$. Hence all
singular values of $D_{\pm}$ are $\sqrt{M}$, and $\trnorm{D_{\pm}} =
M^{3/2}$. Therefore, using triangle inequality 
\begin{equation*}
\trnorm{D} = \trnorm{(D_{\pm} + J)/2} \geq  \frac{1}{2} \left
  (\trnorm{D} - \trnorm{J} \right) = \frac{M^{3/2} - M}{2}.
\end{equation*}
Let $D = EF$ be some optimal decomposition of $D$ for the row-column
width. 
By Holder's inequality, we have 
\begin{equation*}
\trnorm{D} = \trnorm{E F} \leq \frnorm{E} \frnorm{F} \leq \sqrt{M
  \cdot \rn(E)^2} \sqrt{M \cdot \cn(F)^2} = M \cdot \rcw(D)
\end{equation*}
Combining both inequalities 
\begin{equation*}
\rcw(D) \geq \frac{\sqrt{M} - 1}{2}.
\end{equation*}
\end{proof}

\section{Open problem}
The main question left open by this work is whether it is possible
to overcome the exponential overhead incurred in simulating a 
classical public coin simultaneous message protocol by a quantum
simultaneous message protocol with no shared resource. Interesting
progress on this question has been made by the recent paper of
Gavinsky, Kempe and de Wolf~\cite{gkw:sim}.

\section*{Acknowledgements}
We are very grateful to Ronald de Wolf for allowing us to include his
proof of Fact~\ref{fact:ronald} in this note.

\bibliography{fingerprint}

\end{document}